\documentclass[aps, twocolumn, prd, showpacs, preprintnumbers, amsmath, amssymb, nofootinbib]{revtex4-1}

\usepackage{graphicx}%
\usepackage{graphics}
\usepackage{dcolumn}%
\usepackage{bm}%

\begin{document}
\title{Diffuse Galactic gamma-ray emission with H.E.S.S.} 

\author{H.E.S.S. Collaboration} 
\author{A.~Abramowski$^{1}$}
\author{F.~Aharonian$^{2,3,4}$}
\author{F.~Ait Benkhali$^{2}$}
\author{A.G.~Akhperjanian$^{5,4}$}
\author{E.O.~Ang\"uner$^{6}$}
\author{M.~Backes$^{7}$}
\author{S.~Balenderan$^{8}$}
\author{A.~Balzer$^{9}$}
\author{A.~Barnacka$^{10,11}$}
% ...
\vspace{10mm}

\footnotesize
\affiliation{
$^{1}$
Universit\"at Hamburg, Institut f\"ur Experimentalphysik, Luruper Chaussee 149, D 22761 Hamburg, Germany} 
 \affiliation{$^{2}$
Max-Planck-Institut f\"ur Kernphysik, P.O. Box 103980, D 69029 Heidelberg, Germany} 
\affiliation{$^{3}$
Dublin Institute for Advanced Studies, 31 Fitzwilliam Place, Dublin 2, Ireland} 
\affiliation{$^{4}$
National Academy of Sciences of the Republic of Armenia,  Marshall Baghramian Avenue, 24, 0019 Yerevan, Republic of Armenia} 
\affiliation{$^{5}$
Yerevan Physics Institute, 2 Alikhanian Brothers St., 375036 Yerevan, Armenia} 
\affiliation{$^{6}$
Institut f\"ur Physik, Humboldt-Universit\"at zu Berlin, Newtonstr. 15, D 12489 Berlin, Germany} 
\affiliation{$^{7}$
University of Namibia, Department of Physics, Private Bag 13301, Windhoek, Namibia} 
\affiliation{$^{8}$
University of Durham, Department of Physics, South Road, Durham DH1 3LE, U.K.} 
\affiliation{$^{9}$
GRAPPA, Anton Pannekoek Institute for Astronomy, University of Amsterdam,  Science Park 904, 1098 XH Amsterdam, The Netherlands} 
\affiliation{$^{10}$
Obserwatorium Astronomiczne, Uniwersytet Jagiello{\'n}ski, ul. Orla 171, 30-244 Krak{\'o}w, Poland} 
\affiliation{$^{11}$
now at Harvard-Smithsonian Center for Astrophysics,  60 Garden St, MS-20, Cambridge, MA 02138, USA} 
% ...
\begin{abstract}
\normalsize
Diffuse $\gamma$-ray emission is the most prominent observable signature of celestial cosmic-ray interactions at high energies. While already being investigated at GeV energies over several decades, assessments of diffuse $\gamma$-ray emission at TeV energies remain sparse. After completion of the systematic survey of the inner Galaxy, the H.E.S.S. experiment is in a prime position to observe large-scale diffuse emission at TeV energies. Data of the H.E.S.S. Galactic Plane Survey are investigated in regions off known $\gamma$-ray sources. Corresponding $\gamma$-ray flux measurements were made over an extensive grid of celestial locations. Longitudinal and latitudinal profiles of the observed $\gamma$-ray fluxes show characteristic excess emission not attributable to known $\gamma$-ray sources. For the first time large-scale $\gamma$-ray emission along the Galactic Plane using imaging atmospheric Cherenkov telescopes has been observed.
While the background subtraction technique limits the ability to recover modest variation on the scale of the H.E.S.S. field of view or larger, which is characteristic of the inverse Compton scatter-induced Galactic diffuse emission, contributions of neutral pion decay as well as emission from unresolved $\gamma$-ray sources can be recovered in the observed signal to a large fraction. Calculations show that the minimum $\gamma$-ray emission from $\pi^0$-decay represents a significant contribution to the total signal. This detection is interpreted as a mix of diffuse Galactic $\gamma$-ray emission and unresolved sources.
\end{abstract}
\maketitle
\normalsize
\section{Introduction}
Cosmic rays permeate our Galaxy and thereby undergo interactions, producing amongst other particles diffuse $\gamma$-rays at high energies. Interactions capable of producing $\gamma$-rays are the production and subsequent decay of neutral pions in the interstellar medium, inverse Compton scattering on radiation fields and bremsstrahlung. Each of these processes contributes differently depending on energy and line-of-sight integrated densities of matter or radiation fields. Diffuse $\gamma$-ray emission was observed first by SAS-2 \citep{SAS2} and further investigated by COS-B \citep{COSB} and EGRET \citep{EGRET}. The most recent and detailed survey of the $\gamma$-ray sky employed the Fermi-LAT instrument studying the energy range between 200~MeV and 100~GeV \citep{FermiDiffuse}. At these energies, the diffuse emission constitutes the principal component of the $\gamma$-ray sky, and represents emission originating from cosmic-ray interactions, not dominated by unresolved sources \citep{PopSynthesis2007}.\\
Towards higher energies, the $\gamma$-ray flux from resolved sources increasingly dominates the total observed celestial $\gamma$-ray emission. Accordingly, the respective signal at these energies contains a potentially large fraction of unresolved $\gamma$-ray sources. At energies close to $\sim$1~TeV, bremsstrahlung becomes irrelevant and the remaining interaction processes that yield very-high-energy (VHE, $E>100$~GeV) $\gamma$-rays are inverse Compton scattering and neutral pion decay \cite{FelixBook}. In this energy regime, observations of diffuse $\gamma$-ray emission have been reported by the Milagro experiment \citep{MilagroDiffuse} at a median energy of 15~TeV, and also by ARGO-YBJ \citep{ARGO}. These experiments operate under favourable duty cycles and observe large fields of the sky, yet they are limited in $\gamma$-hadron separation quality as well as angular resolution, which further complicates conclusive discrimination between $\gamma$-ray sources and diffuse emission signatures.
Profiting from a substantially lower energy threshold and arc-minute scale angular resolution, imaging atmospheric Cherenkov telescopes have the potential to improve substantially on these measurements. Particularly the High Energy Stereoscopic System (H.E.S.S.) is privileged due to the comparatively large field of view ($5^\circ$ in diameter) and its location in Namibia, which allows for an excellent view on the central part of the Galactic Plane. \\
Presented here is a first study of the diffuse $\gamma$-ray emission utilizing the imaging atmospheric Cherenkov technique. Problems arising in this measurement and methodological limitations imposed by the technique are discussed. The resulting signal is interpreted with respect to cosmic-ray interactions via neutral pion decay, inverse Compton scattering, and contributions of unresolved sources.
\section{Data and Analysis Methodology}
\subsection{The H.E.S.S. Galactic Plane Survey}
H.E.S.S. is a system of imaging atmospheric Cherenkov telescopes in the Khomas highland of Namibia \cite{Crab2006}. A system of four telescopes has been taking data since 2003. Since 2012 H.E.S.S. advanced to its second phase featuring a central telescope with a sixfold mirror area compared to the original 12~m diameter telescopes.\\
A substantial part of the H.E.S.S. I data set is the H.E.S.S. Galactic Plane Survey (HGPS), which was accumulated over the past 10 years with the four telescope system.
Although having revealed a wealth of new sources \citep{HGPSICRC2013}, and already reported extended emission from the Galactic Center ridge \citep{Diffuse2006}, a measurement of the large-scale diffuse $\gamma$-ray emission remains challenging regarding sensitivity and analysis methodology.
\subsection{Analysis of the Galactic Plane Survey}
Our investigation of diffuse emission relates to those regions where no $\gamma$-ray sources are detected. \\
As such an analysis aims at discovery of a very faint
signal, the application of a sensitive analysis method is required.
The results presented here are obtained using semi-analytical modeling of the air shower produced in $\gamma$-ray interactions in the atmosphere, resulting in an improved sensitivity compared to conventional analysis methods \citep{Model}. Good-quality data with pointings in the region between $-77.5^\circ$ and $62.5^\circ$ in Galactic longitude $l$ and between $-4.5^\circ$ and $4.5^\circ$ in Galactic latitude $b$ are used for a measurement in the range of $-75^\circ < l < 60^\circ$ and $-2^\circ < b < 2^\circ$, where the exposure of the HGPS is largest. The data amount to a total of $2484.6$ hours of dead-time corrected observation time. The overall average of the energy threshold of the analysis is roughly 250~GeV.
Standard analysis cuts \citep{Model} reduce the background and guarantee good quality of the event reconstruction. Results have been cross-checked with an independent calibration and analysis procedure with a boosted-decision-tree-based Hillas parameter technique \cite{TMVA} for consistency.
\subsection{Background Subtraction}
Hadrons and electrons can produce air showers that look like $\gamma$-ray showers and it is therefore necessary to subtract remaining
background events that survive event selection cuts. The standard procedure to avoid systematic effects caused by changing atmospheric or instrument conditions, is to determine the background level from data, from the same field of view, in regions of no known $\gamma$-ray sources \citep{BGBerge}. However, this method places constraints on the size of the emission region that can be probed, which has to be smaller than the field-of-view to allow for background estimation regions.
Special attention needs to be paid to the selection of celestial regions applicable to a background measurement.
In order to define the regions excluded from background subtraction, an iterative procedure is adopted. At each step, a significance map of the Galactic Plane region is computed using the ring background technique \cite{HGPSICRC2013} with an oversampling radius of $0.22^\circ$ (suitable for slightly extended sources). The following exclusion conditions apply: Each pixel\footnote{The pixel size in the maps is $0.02^\circ \times 0.02^\circ$.} with a significance $s$ above 4 $\sigma$ with at least one neighboring pixel with $s > 4.5 \sigma$ is excluded and vice versa. In order to include also tails in the PSF used to describe the $\gamma$-ray sources, the obtained exclusion regions are extended by $0.2^\circ$. This procedure is repeated until the significance distribution of the non-excluded pixels has a normal shape with $|\mu| < 0.05$ and $w < 1.1$ ($\mu$ and $w$ being the mean and the width of the distribution respectively). The resulting excluded regions are visualized by the dark areas in Fig.~\ref{fig1}. In addition, the complete region along the Galactic Plane with a latitude range of $-1.2^\circ < b < 1.2^\circ$ is excluded (visualized by the horizontal dashed lines in Fig.~\ref{fig1}).
The choice of the latitude range is a compromise between a desired large excluded region in order to avoid contamination of the background estimate on the one hand and the need for statistics and reduction of systematics in the background measurement on the other hand. An adaptive ring background subtraction method has been chosen \citep{HGPSICRC2013} to allow for optimal choices of background regions.\\
A consequence of the applied background subtraction is that the method used is rather insensitive to large-scale emission with modest variation in latitudinal intensity because such signals are subtracted along with the background.
The observed signal therefore needs to be interpreted as excess relative to the $\gamma$-ray emission at absolute latitudes exceeding $|b| = 1.2^\circ$.
\subsection{Generation of flux maps}
For the region of $-75^\circ<l<60^\circ$ and $-2^\circ<b<2^\circ$ a map of the differential flux normalization at 1~TeV is obtained from the background-subtracted $\gamma$-ray excess map by division by the integrated exposure map:
$\phi = n_\gamma / \sum A_\mathrm{int} t_\mathrm{obs}$.
The exposure is summed over individual observation positions, with integrated acceptance $A_\mathrm{int}$ and dead-time corrected observation time $t_\mathrm{obs}$. The integrated acceptance is obtained from simulations and requires a spectral assumption, which is a powerlaw with spectral index of $2.2$.
The result turns out to be only weakly sensitive to the choice of spectral index (with deviations in regions off known $\gamma$-ray sources of less than $5\%$ when altering the spectral index assumption to $2.7$).
\subsection{Definition of the Analysis Regions}
\begin{figure*}
\begin{center}
   \includegraphics[width=18cm]{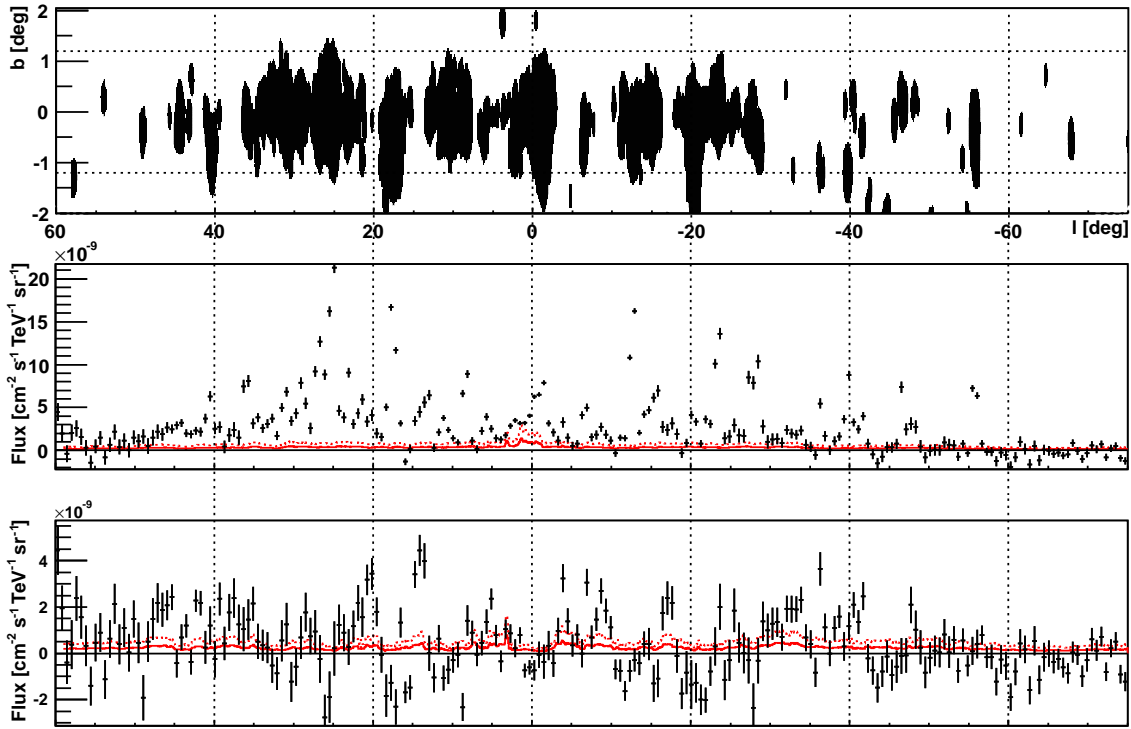}
\end{center}
     \caption{Top panel: The white regions depict the diffuse analysis region (DAR). Black are regions of significant $\gamma$-ray emission. Horizontal dashed lines mark the region $-1.2^\circ<b<1.2^\circ$ that is excluded from background subtraction. Middle panel: The longitudinal profile of the Galactic Plane over a latitude range of $-2^\circ<b<2^\circ$. Shown is the differential flux at 1~TeV including sources. H.E.S.S. TeV data, which include known sources, are indicated by black crosses. The minimal 1~TeV $\gamma$-ray emission from hadronic interactions, estimated using HI and H$_2$ data (traced by CO data) and a solar-like cosmic-ray spectrum (see text), is shown as model curve. The dashed line includes a nuclear enhancement factor of 2.1. Model curves do not comprise a reduction due to background subtraction. Bottom panel: The same as the middle panel, except only the DAR is considered. The distribution is strongly influenced by the shape of the DAR ({\it cf.} top panel). Model curves correspond to the minimal hadronic $\gamma$-ray emission expected in the same region.}
   \label{fig1}
\end{figure*}

In the following sections total flux distributions are compared with those of regions that do not contain significantly detected $\gamma$-ray sources. These regions are labelled diffuse analysis region (DAR) and are defined in the same way as regions suitable for background measurements. The DAR is shown in Fig.~\ref{fig1}.\\
As the Galactic Plane contains a large number of extended sources (including those with complex morphology), the percentage of regions excluded from the DAR amounts to $20 \%$, whereas in the latitudinal region $-0.5^\circ < b < 0.5^\circ$ this percentage increases to more than $40 \%$. 
\subsection{Profile Generation}
For an investigation of the distribution of $\gamma$-ray flux profiles in Galactic longitude and latitude are generated. These profiles are obtained by integrating the flux map over either longitude or latitude and by normalizing to the covered area, thus resulting in an average flux profile for the latitudinal and longitudinal region considered. This procedure is done once for the complete data set and once for the DAR. The resulting profiles including $1 \sigma$ uncertainties are shown in Figs.~\ref{fig1} and \ref{fig2}.
\section{Results}
\subsection{Spatial characteristics of the signal}
\begin{figure}
\centering
   \includegraphics[width=9cm]{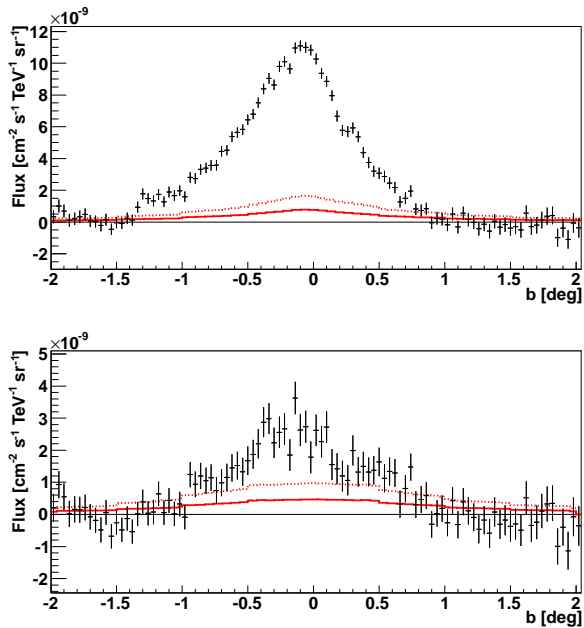}
     \caption{Top panel: The latitudinal profile of the Galactic Plane over a longitude range of $-75^\circ<l<60^\circ$. Shown is the differential flux at 1~TeV including sources. H.E.S.S. TeV data, which include known sources, are indicated by black crosses. The minimal 1~TeV $\gamma$-ray from hadronic interactions, estimated using HI and H$_2$ data (traced by CO data) and a solar-like cosmic-ray spectrum (see text), is shown as model curve. The dashed line includes a nuclear enhancement factor of 2.1. Model curves do not comprise a reduction due to background subtraction. Bottom panel: The same as the top panel, except only the DAR (for the definition see Fig.~\ref{fig1} top panel) is considered. Model curves correspond to the minimal hadronic $\gamma$-ray emission expected in the same region.}
     \label{fig2}
\end{figure}

The longitudinal profile in Fig.~\ref{fig1} shows a spiky distribution of the Galactic $\gamma$-ray sources for the complete region (middle panel). For the DAR (bottom panel) fluxes are on average positive (although hardly significant in most individual bins). For the signal, a clear correlation with the distribution of the excluded regions in the DAR can be seen: excess is observed only in longitude ranges with sparse exclusion of regions at small latitudes.
Zero or even mildly negative fluxes are found when large regions close to the Galactic Equator are excluded from the DAR. The reason for this is an over-subtraction of the background determined from signal-contaminated regions.
The reflection of the shape of the DAR in the longitudinal profile strongly limits its potential in terms of a physics interpretation. However, it can be seen that the signal does not originate from left-over contributions of excluded sources but it rather accumulates over longitude.\\ 
The latitudinal profiles of both the complete data set and the DAR, shown in Fig.~\ref{fig2}, exhibit a clear excess over zero. The significance of the detected signal has been evaluated by comparing the observed latitudinal profile to a zero-flux baseline hypothesis as a function of Galactic latitude. Whereas the full latitudinal profile ($|b| \le 2^\circ$) has been found to deviate by more than 6 $\sigma$ from the null hypothesis, the significance increases to above 20 $\sigma$ in those latitudinal ranges that are close to the Galactic Equator.\\ 
The latitudinal profile exhibits a maximum of around $3\times 10^{-9}$~TeV$^{-1}$~s$^{-1}$~cm$^{-2}$~sr$^{-1}$, at a latitude slightly shifted from the Galactic Equator towards negative values ($b_\mathrm{max}\approx -0.2^\circ$). This is similarly observed for the total flux, which contains all $\gamma$-ray sources in addition to the diffuse emission. The distribution falls off towards higher latitude values and reaches zero flux at latitude of $b \approx \pm 1^\circ$. As a consequence of the applied background subtraction, slightly negative fluxes can be observed at $b \approx \pm 1.5^\circ$ (see previous discussion).
In comparison with the total flux, the signal in the DAR makes up $\sim 28 \%$ in the central $2^\circ$ region. A large-scale signal prevails in the longitudinal profile regardless of the details in the definition of the DAR.\\
Systematic uncertainties enter at several stages in the presented analysis. A comparison with an independent cross-check analysis with separate calibration procedures of the data resulted in consistency within $\sim 30\%$ in flux normalisation. This uncertainty, however, does not account for the effect of a reduced signal due to the applied background subtraction, which is present in both analysis chains.
The influence of the background removal technique can be determined under a model assumption for the $\gamma$-ray emission: a Gaussian of width $2^\circ$ in latitude results in a reduction of $30 \%$ of the original signal, a Gaussian of width $20^\circ$ in a reduction of $95\%$.

\subsection{Assessment of the Detection}
The observed signal can originate from hadronic emission of cosmic-ray interactions with matter via $\pi^0$ production and decay, inverse Compton scattering of cosmic-ray electrons off radiation fields and unresolved $\gamma$-ray sources. The contributions of these possible origins are discussed in the following sections.
\subsubsection{Hadronic emission}
The component of hadronic emission is constrained by the level of cosmic rays and the total target material. For an estimation of the minimum, guaranteed contribution to be present in our observed signal, the emission from the sea of cosmic rays (assumed to resemble the locally measured cosmic-ray spectrum) interacting with gas content (indicated from respective spectral line observations) is calculated. This minimum $\gamma$-ray emission related to hadronic gas interactions undergoes the same spatial selection as H.E.S.S. flux maps for the production of profiles. Results of these calculations are shown in Figs.~\ref{fig1} and \ref{fig2}, together with the H.E.S.S. data, as red model curves.\\ 
Gas templates of HI and H$_2$ column densities are used for the calculation: HI data originate from the Leiden/Argentine/Bonn Survey \citep{LAB}, a column density is obtained assuming a spin temperature of $T_S = 125$~K. The H$_2$ column density is traced by CO (1-0) measured by the NANTEN telescope. The conversion factor is chosen to be $X_\mathrm{CO} = 2 \cdot10^{20}$~cm$^{-2}$~K$^{-1}$~km$^{-1}$~s \citep{XCO}. Since the degeneracy between HI, H$_2$ and dust-related tracers for energetic $\gamma$-ray emission is not yet satisfactorily resolved at lower energies - where the majority of all observed photons is attributed to diffuse Galactic emission \citep{2012ApJ...755...22A, 2011ApJ...726...81A} - an additional dust-related (dark gas) component is not considered here.\\
The minimum expected $\gamma$-ray flux is obtained from integrating the product of the gas column density $n(l,b)$,
the interaction cross section $\frac{d\sigma_{CR\longrightarrow \gamma}}{dE_{CR}}$, and the cosmic-ray energy spectrum $J(E_{CR})$ \citep{PDG} over energy: 
\begin{equation*}
\frac{dF(l,b)}{dE_\gamma} = \int \frac{d\sigma_{CR\longrightarrow \gamma}}{dE_{CR}}\, n(l,b)\, J(E_{CR})\,dE_{CR}\,\,.
\end{equation*} 
\\
The parametrization of the interaction cross section follows Kelner et al. \cite{Kelner}. H$_2$ is treated as two individual protons. For a conservative minimum in the calculated $\gamma$-ray emission, the proton cross section is applied also for heavier cosmic-ray nuclei. A nuclear enhancement factor accounting for contributions of nucleonic cosmic-ray interactions (beyond proton-proton) to the diffuse $\gamma$-ray emission is model-dependent but typically considered in the range of 1.5 to 2 (see \cite{Mori} and references therein). 
In Figs.~\ref{fig1} and \ref{fig2} the corresponding flux according to a more recent estimate of $\approx 2.1$ by Kachelriess et al. \cite{Kachelriess2014} is indicated by a dashed line.\\
When comparing the shape of the distributions, a difference can be observed in the widths of the latitudinal profiles: The hadronic component exhibits a FWHM of $2^\circ$. The H.E.S.S. data exhibits a narrower width of $1^\circ$ for the total flux including $\gamma$-ray sources, while the profile of the DAR has a FWHM of $1.2^\circ$ - slightly broader, which could hint at a composite origin of the DAR signal, consisting of both $\gamma$-ray sources and hadronic diffuse emission. 
Considering the fraction of the hadronic contribution, the minimum estimated from p-gas interactions in the range of $-1^\circ < b < 1^\circ$ is $9\%$ for the total flux and $26 \%$ for the DAR. These values increase to $19\%$ (total) and $55 \%$ (DAR) when considering the nuclear enhancement factor. The background subtraction that is applied to the H.E.S.S. data reduces the detectable $\gamma$-ray emission by around a third, yielding fractions of $14\%$ (total) and $36 \%$ (DAR) for the hadronic contribution in the respective signal.
\subsubsection{Large-scale inverse Compton emission}
Another major contribution to the diffuse emission signal at very high energies is predictably related to continuous cosmic-ray electron and positron energy losses via inverse Compton (IC) scattering. Both existence and relevance of an IC-emission contributing to an observable diffuse emission signal can be deduced from the immediately neighboring energy band, the Galactic diffuse $\gamma$-ray emission at GeV energies. Studies of the Galactic diffuse emission in the Fermi-LAT energy range \citep{FermiDiffuse} indicated contributions by IC-scattering to the total observed diffuse emission with an intensity up to the same order of the pionic emission component. More specifically, IC-related $\gamma$-ray emission was reported at similar intensity to the hadronic $\gamma$-ray emission produced from gas traced by HI for high Galactic latitudes, and dominant above tens of GeV \citep{FermiDiffuse}. Spectral extrapolation is suggestive of both hadronic and IC-related emission components extending towards even higher energies before either energy losses soften or even cut-off the IC-spectrum, or the neutral pion production spectrum might indicate the imprint of the maximum energy reached by particle acceleration in our Galaxy.
At first glance, the IC-emission component used to interpret the Fermi-LAT detected diffuse Galactic emission might serve as a reasonable template for such an extrapolation. Respective predictions were derived on the basis of a 2D slab model by the GALPROP\footnote{http://galprop.sourceforge.net/, http://galprop.stanford.edu} cosmic-ray propagation code and an interstellar radiation field model \citep{FermiDiffuse} available alongside. The intensity distribution is generally smooth, with mild gradients along the Galactic Plane and comparably steep gradients towards higher Galactic latitudes. The latitudinal intensity profiles are significantly more extended compared to those derived from the gas template. The present analysis framework allows only for partial recovery of such gradients. For the case of a GALPROP-based prediction, the aforementioned background subtraction would yield a reduction of $\sim 95\%$ of the celestial IC flux.\\
More realistic predictions for an IC-component at TeV energies are not expected to resemble the smoothness from lower GeV energies since energy losses increasingly confine the emission to local sources or source regions. Accordingly, an imprint from sources as well as Galactic structure is anticipated \citep{FermiDiffuse}. Developments beyond limitations of present diffuse data interpretation and modelling (e.g. as indicated in \cite{FermiDiffuse} and \cite{2013ApJ...777..149V}) are ongoing. 
\subsubsection{Contribution from unresolved sources}
A third major contribution to the detected high GeV to low TeV signal is related to the existence of VHE $\gamma$-ray sources below instrumental detection threshold, namely unresolved sources. The H.E.S.S. Galactic survey region is clearly dominated by emission of individual sources \citep{HGPSICRC2013}, and there is every reason to assume that this source wealth continues below the current H.E.S.S. detection threshold. The sensitivity accomplished with the HGPS does not comprise the depth of the whole Milky Way ({\it cf.} Fig.~4 in \cite{HGPSICRC2013}), accordingly unresolved sources will contribute to the large-scale emission signal that is discussed here. This is not an unexpected situation. Contributions from unresolved sources are expected to increasingly contribute to the detected emission signal towards higher energies. Also, the $\gamma$-ray flux from unresolved sources will not suffer a heavy suppression from the background removal, since the population of unresolved $\gamma$-ray sources is likely to follow the distribution of the resolved sources, narrowly localized along the Galactic Plane.\\
Refined analysis of the number of detected sources vs. cumulative flux distribution (log N($>$S) - log S) of the Galactic VHE source population \cite{2009Matthieu} on the basis of the upcoming H.E.S.S. legacy source catalog project \cite{HGPSICRC2013} will allow for a quantitative assessment of the contribution of unresolved sources to the observed signal. \\
\section{Conclusion}
This paper presents the first detection of large-scale $\gamma$-ray emission along the Galactic Plane using imaging atmospheric Cherenkov telescopes. A significant flux along the Galactic Plane is detected, which is not attributed to resolved and significantly detected $\gamma$-ray sources. The detection can be interpreted as diffuse Galactic $\gamma$-ray emission and contributions from unresolved sources. 
Owing to limitations of the applied background removal technique, modest variations in the emission on the scale of the H.E.S.S. field of view are suppressed in such measurements. As a consequence, the reported signal is considered to represent a lower limit compared to what might be detected with improved analysis strategies at these energies.\\
The observed signal is comprised of contributions from cosmic-ray interactions via neutral pion decay-induced $\gamma$-ray emission and inverse Compton scattering as well as from unresolved $\gamma$-ray sources. The flux of the $\gamma$-ray emission related to $\pi^0$ decay is estimated via line-of-sight column densities in HI and CO with the corresponding narrow latitudinal profile. Such low-scale-height components are not severely impacted by the applied background subtraction. The same can be expected for the contribution from unresolved sources. In contrast, inverse Compton emission is expected to have a distinctly larger scale height and is, as such, only partly recoverable. While a guaranteed contribution of $\gamma$-ray emission from cosmic-ray interactions with the interstellar medium already makes up a sizable fraction of the signal, the nature of the remaining excess flux and its division among the different emission components remains to be conclusively identified.

\begin{acknowledgments}
The support of the Namibian authorities and of the University of Namibia
in facilitating the construction and operation of H.E.S.S. is gratefully
acknowledged, as is the support by the German Ministry for Education and
Research (BMBF), the Max Planck Society, the German Research Foundation (DFG),
the French Ministry for Research,
the CNRS-IN2P3 and the Astroparticle Interdisciplinary Programme of the
CNRS, the U.K. Science and Technology Facilities Council (STFC),
the IPNP of the Charles University, the Czech Science Foundation, the Polish
Ministry of Science and  Higher Education, the South African Department of
Science and Technology and National Research Foundation, and by the
University of Namibia. We appreciate the excellent work of the technical
support staff in Berlin, Durham, Hamburg, Heidelberg, Palaiseau, Paris,
Saclay, and in Namibia in the construction and operation of the
equipment.
\end{acknowledgments}

\end{document}